\documentclass[preprint,aps,amsmath,amssymb,eqsecnum]{revtex4}
\usepackage{graphicx}
\usepackage{wrapfig}
\usepackage{subfigure}
\usepackage{mathptmx}
\usepackage{psfig}
\usepackage{dcolumn}
\usepackage{bm}
\usepackage{tabls}
\newcommand{\be}{\begin{eqnarray}}
\newcommand{\ee}{\end{eqnarray}}
\begin{document}
\title{A transition in  the spectrum of the topological sector 
of $\phi_2^4$ theory at strong coupling}
\author{Dipankar Chakrabarti}
\email{dipankar@phys.ufl.edu}
\affiliation{Physics Department, University of Florida, Gainesville, FL
32611, U.S.A.}
\author{A. Harindranath}
\email{hari@theory.saha.ernet.in}
\affiliation{Theory Group, Saha Institute of Nuclear Physics \\
 1/AF Bidhan Nagar, Kolkata 700064, India}
\author{J. P. Vary}
\email{jvary@iastate.edu}
\affiliation{Department of Physics and Astronomy, Iowa State University,
Ames, IA 50011, U.S.A.}

\date {April 8, 2005}

\begin{abstract}

We investigate   the strong coupling region of the topological sector
 of  the two-dimensional $\phi^4$ theory. Using  discrete light cone 
quantization (DLCQ), we extract the masses of the lowest few excitations 
and observe level crossings. To understand this phenomena, we evaluate 
the expectation value of the integral of the normal ordered $\phi^2$ 
operator and we extract  the number  density of constituents in these 
states.  A  coherent state variational calculation confirms that the
number density for low-lying states above the transition coupling
is dominantly that of a kink-antikink-kink state. The Fourier transform 
of the form factor of the lowest excitation is extracted which
reveals a structure close to a kink-antikink-kink profile.  Thus, we 
demonstrate  that the  structure of the lowest excitations becomes that 
of  a kink-antikink-kink configuration at moderately strong coupling. We 
extract the critical coupling for the transition of the lowest state 
from that of a kink to a kink-antikink-kink. We interpret the transition
as evidence for the onset of kink condensation which is believed to be   
the physical mechanism for the symmetry restoring phase transition in 
two-dimensional $\phi^4$ theory.

\end{abstract}
\maketitle
\section{Introduction}

As is well known \cite{dashen, rajaraman, gj}, topological excitations (kinks) 
exist in 
classical two-dimensional $\phi^4$ model
with negative quadratic term (broken phase).
However, the study of topological objects in quantum field theory is highly 
nontrivial. Most of the investigations to date use semi-classical or
Hartree approximations.
Using the techniques of constructive quantum field theory \cite{glimmj}, 
it was proven rigorously that in quantum theory, a stable kink state is
separated from the vacuum by a mass gap of the order $\lambda^{-1}$ and
from the rest of the spectrum by an upper gap \cite{froh}. More detailed
nonperturbative information on the spectrum of the mass operator or on other
observables from rigorous approaches is not available.
It is worthwhile to recall
that the study of these objects in lattice field theory is also highly
non-trivial \cite{Ciria,Ardekani}. 
For very recent work on the kinks in
two dimensional $\phi^4$ theory in the Hartree approximation,
see Refs. \cite{bb,salle}.

Light front quantization offers many advantages for the study of two
dimensional quantum field theories. As is well-known, of the three
Poincare generators  only the Hamiltonian is dynamical. In contrast,
in the conventional instant form formulation, only momentum
is a kinematical operator. Kinematical boost invariance allows for the
extraction of Lorentz invariant observables such as the parton distribution
function which has the simple interpretation of a number
density. The formulation allows for the extraction of the spectrum
of the invariant mass operator in a straight-forward manner. In this work
we utilize the discrete version of the light front quantization, namely,
discrete light cone quantization (DLCQ) \cite{dlcq} for numerical
investigations of two dimensional $\phi^4$ theory with anti periodic
boundary condition (APBC) at strong coupling. 
Since the zero momentum mode
is absent, the Hamiltonian has the simplest Fock space structure in this
case.

The quantum kink on the light front was first addressed by Baacke
\cite{Baacke} in the context of semi-classical quantization.
Recently Rozowsky and Thorn \cite{RT}, with the help of a coherent state
variational calculation, have shown that it is possible to 
extract the
mass of topological excitations in DLCQ with periodic
boundary condition (PBC) while dropping the zero momentum mode. 
Motivated by the remarkable work of  these authors,
we have initiated  the study \cite{kinkplb} of 
topological objects 
in the broken symmetry phase of
two dimensional $\phi^4$ theory  
using  APBC in DLCQ. 
We presented evidence for
degenerate ground states, which is both a
signature  of spontaneous symmetry breaking and mandatory for the
existence of kinks. Guided by a constrained variational calculation
with a coherent state ansatz, we then extracted the vacuum  energy
and kink mass and compared with classical and semi-classical results.
We compared the DLCQ results for the number density of partons in the kink
state and the Fourier transform of the form factor of the kink  with
corresponding observables in the  coherent variational kink state. 
We have also carried out similar  investigations using PBC \cite{pp1}.

In this work, we probe the strong coupling region of the topological
sector of the theory with topological charge $ \pm 1$. 
The plan of this work is as follows. 
We follow the notations and conventions of Ref. \cite{kinkplb}. 
Major aspects of Hamiltonian diagonalization are given 
in Sec. II.  In Sec. III we extract of the mass of the lowest few
excitations as a function of the coupling and display results for the gaps
in mass-squared. As the coupling increases, we
observe level crossing in the spectrum. In this section, we also discuss the 
manifestation of level crossing in another observable, namely, the expectation
value of the integral of the normal ordered $\phi^2$ operator. 
To gain further insights, we need to
probe other physical properties of the low-lying excitations. Toward this
end, we perform
DLCQ calculations of the parton density and the 
Fourier transform of the form factor of the lowest excitation at  
moderately strong couplings and the results are presented   
in Sec. IV. In the same section a 
coherent state variational calculation of the kink-antikink-kink profile
and corresponding parton density are also  given. 
Sec. V contains a discussion, summary and conclusions. 
\section{Hamiltonian and Diagonalization}

We start from the Lagrangian density
\be
{\cal L} = \frac{1}{2} \partial^\mu \phi \partial_\mu \phi +
 \frac{1}{2} \mu^2 \phi^2 -  \frac{\lambda}{4!} \phi^4.
\ee
The light front variables are defined by $ x^\pm = x^0 \pm x^1$.

The Hamiltonian density
\be
{\cal P}^- =
 - \frac{1}{2} \mu^2 \phi^2 +  \frac{\lambda}{4!} \phi^4
\ee
defines the Hamiltonian
\be
P^- & = & \int dx^- {\cal P}^-~
     \equiv  \frac{L}{2 \pi} H
\ee
where $L$ defines our compact domain $ - L \le x^- \le +L$. Throughout this
work we address the energy spectrum of $H$.

The longitudinal momentum operator is
\be
P^+ & = & \frac{1}{2} \int_{-L}^{+L} dx^- \partial^+ \phi \partial^+ \phi
 \equiv  \frac{2 \pi}{L}K
\ee
where $K$ is the dimensionless longitudinal momentum operator. The mass
squared operator $M^2 = P^+ P^- = KH$.

In DLCQ with APBC, the field expansion has the form
\be
\Phi(x^-) = \frac{1}{\sqrt{4 \pi}} \sum_n \frac{1}{\sqrt{n}}
\left [a_n e^{-i \frac{n \pi}{L} x^-} + a_n^\dagger e^{i \frac{n
\pi}{L} x^-} \right ].
\ee
Here $ n = \frac{1}{2}, \frac{3}{2},\ldots. $


The normal ordered Hamiltonian  is given by

\be
H & = & - \mu^2 \sum_n \frac{1}{n} a_n^\dagger a_n
+ \frac{\lambda}{4 \pi} \sum_{k \le l, m\le n}~ \frac{1}{N_{kl}^2}
~ \frac{1}{N_{mn}^2}~
\frac{1}{\sqrt{klmn}}
 a_k^\dagger a_l^\dagger a_n a_m
\delta_{k+l, m+n} \nonumber \\&~& + \frac{\lambda}{4 \pi} \sum_{k, l \le
m\le n}~ \frac{1}{N_{lmn}^2}~ 
\frac{1}{\sqrt{klmn}}~
  \left [
a_k^\dagger a_l a_m a_n + a_n^\dagger a_m^\dagger a_l^\dagger a_k \right ]~
\delta_{k, l+m+n}
\ee
with
\be N_{lmn} & = & 1 ,~ l \ne m \ne n, \nonumber \\
            & = & \sqrt{2!},~ l=m \ne n, ~ l \ne m=n,\nonumber \\
            & = & \sqrt{3!},~ l=m=n,
\ee
and
\be
N_{kl} & = & 1, k \ne l,~ \nonumber \\
       & = & \sqrt{2!},~ k=l.
\ee

Since the Hamiltonian exhibits the $ \phi \rightarrow - \phi$ symmetry, the
even and odd particle sectors of the theory are decoupled.  When the
coefficient of $\phi^2$ in the Hamiltonian is positive,
 at weak coupling, the lowest state in the odd particle sector is a
single particle carrying all the momentum. In the even particle sector, the lowest
state consists of two particles having equal momentum. Thus for massive
particles, there is a distinct mass gap between odd and even particle
sectors. When the coefficient of $\phi^2$ in the Hamiltonian is negative, 
at weak coupling, 
the situation is drastically different.
Now, the lowest states in the odd and even particle sectors consist of
the maximum number of particles carrying the lowest allowed momentum. Thus, in
the continuum limit, the possibility arises that the states in the 
even and odd
particle sectors become degenerate. In this case, 
  for any state $|e>$ in the even sector
 we find a state $|o>$ in the odd particle sector with the same mass.
 Hence, we can construct two degenerate states of  mixed symmetry which are
 eigenstates of the Hamiltonian:
$     | \alpha> = |e>+|o> $
 and $ |-\alpha> = |e>-|o>$,
 Under $\phi$  parity $| \alpha> ~ \rightarrow ~|-\alpha>$ and $ |-\alpha> 
~\rightarrow | ~\alpha>$. These
eigenstates
 of the Hamiltonian do not respect the symmetry of the Hamiltonian and
hence
a clear signal of SSB is the degeneracy of the
spectrum in the even and odd particle sectors. 
Recall that with APBC, for integer (half integer) 
$K$, we have even (odd) particle sectors.   Thus at finite $K$,  
we can compare the spectra for an integer K value and its neighboring half
integer $K$ value and look for 
degenerate states. 

All the results presented here were obtained on clusters of computers 
($ \le $ 15 processors) using the many fermion dynamics (MFD) code 
adapted to bosons \cite{mfd}. The Lanczos diagonalization 
method is used in a highly scalable
algorithm that allows us to proceed to high enough values 
of $K$ for smooth $ K \rightarrow \infty $ extrapolations. For our largest
value, $K=60$, the Hamiltonian matrix reaches the dimensionality 1,928,175,
 with configurations up to 120 bosons participating.  
\section{Level crossing in the spectrum at strong coupling}
In Ref. \cite{kinkplb} we presented the lowest four eigenvalues as a
function of $\frac{1}{K}$ for $ \lambda=1.0$ and the extracted values of the
vacuum energy density and kink mass.  We saw that the values extracted
are close to the semi-classical result.

\begin{figure}[hbtp]
\centering
\subfigure[]{
\includegraphics[width=4.2in,clip]{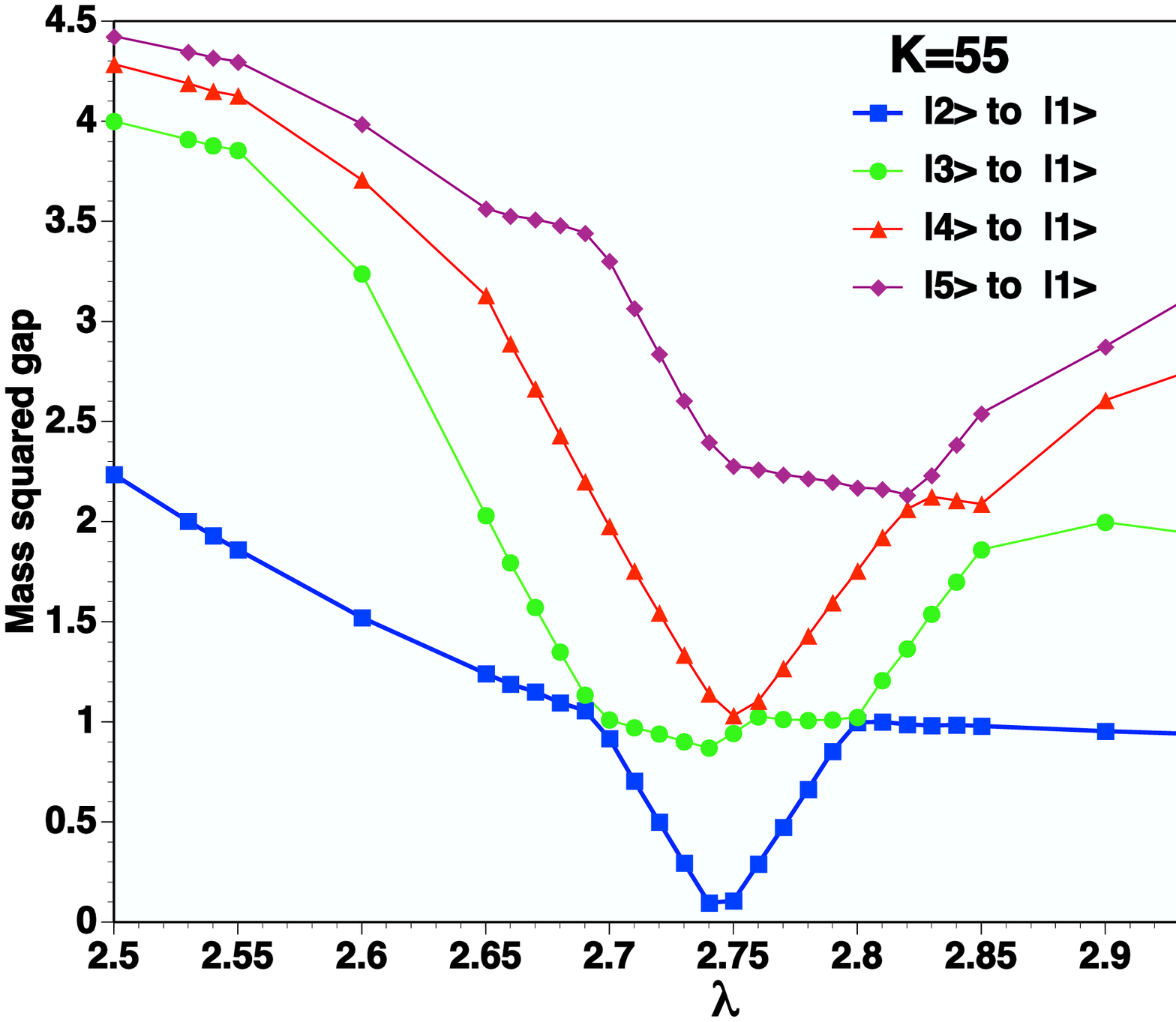}}
\\[1pt]
\subfigure[]{
\includegraphics[width=4.2in,clip]{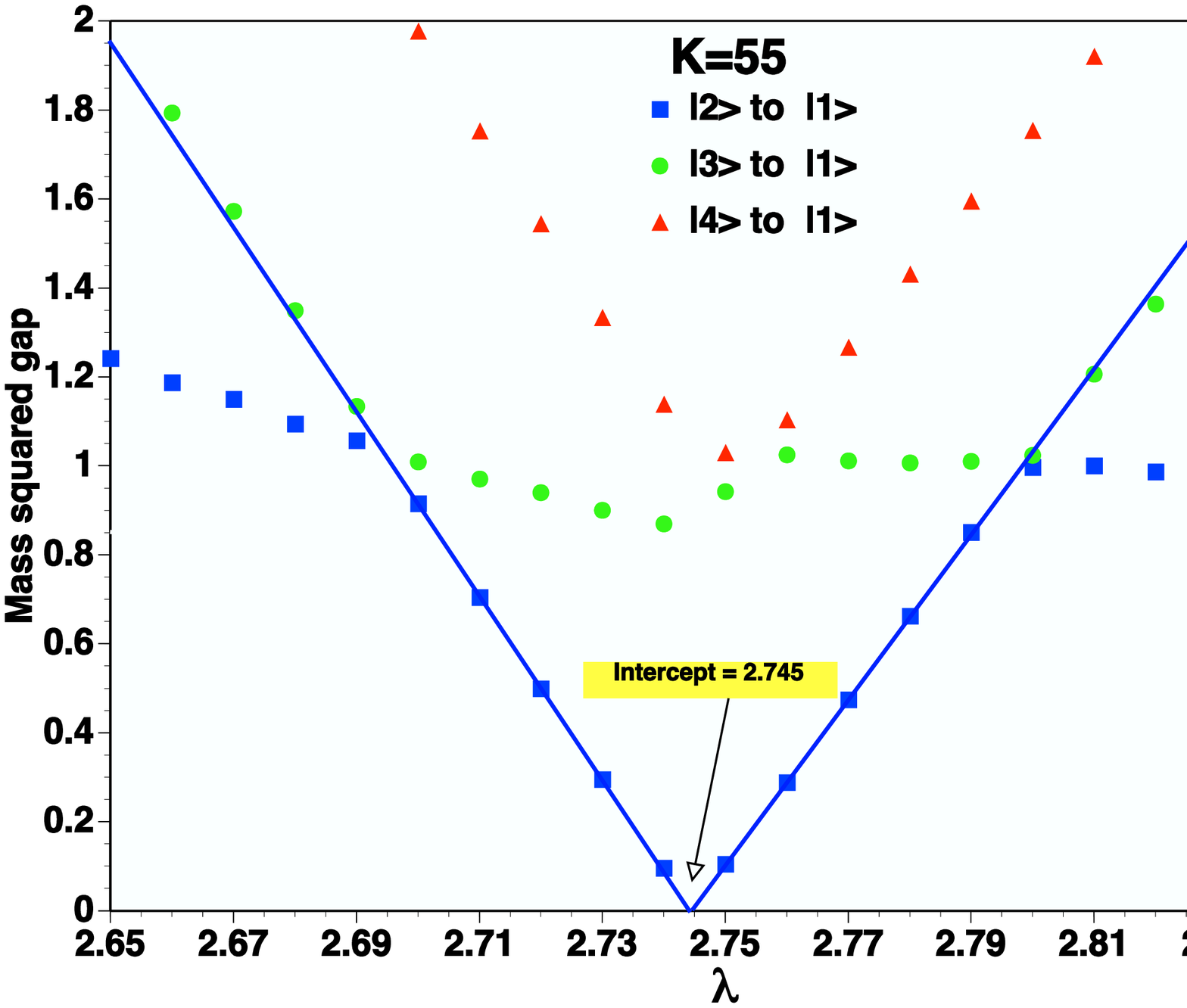}}
    \caption{(Color online)(a) Mass$^{2}$ gap 
as a function of $\lambda$ for $K$=55.  All
calculated results are connected by straight line segments to guide the eye.
(b) Same as in (a) but the
detailed structure for a narrow range of $\lambda$ around the critical
coupling is shown. Here,  two straight lines have been fit to the
lowest gap near the crossing point to extract the $``$critical" value
of the coupling as indicated in the figure.}
\label{mass2gapfig1}
\end{figure}

As we increase the coupling, we find that the lowest two energy levels 
cross each other. The value of the coupling at which this occurs 
lowers as $K$ increases. In Fig. \ref{mass2gapfig1}(a),
 we present the mass squared
differences for the lowest five excitations as a function of $\lambda$ 
in the critical region for the
even ($K=55$) particle sector.  
Many interesting features are exhibited in this figure. Not only the
lowest two levels but all the low-lying levels cross. Before the
occurrence of level crossing, the nature of the spectrum is as follows. 
There is a substantial size to the mass-squared 
gap between the lowest 
three excitations. Thereafter, the gaps become more closely packed, and
more evenly spread.

In the semi-classical analysis \cite{rajaraman}, the
lowest states, in order of excitation,  are kink, excited kink, kink plus
boson, and the continuum states. The wide gaps that exist between the
second and first level and the third and first level and the close packing
of higher levels  are consistent with the expectations from
semi-classical analysis. But as we see in Fig. \ref{mass2gapfig1}(a), 
the gaps vanish at strong
coupling. After the level crossings, we observe almost equally spaced gaps 
for the mass-squared of {\em all} the low-lying excitations
suggestive of a harmonic mass-squared spectrum.
This indicates that the 
nature of low lying levels are substantially different in the two regions
of the coupling.   

\begin{figure}[hbtp]
\centering
\includegraphics[width=5in,clip]{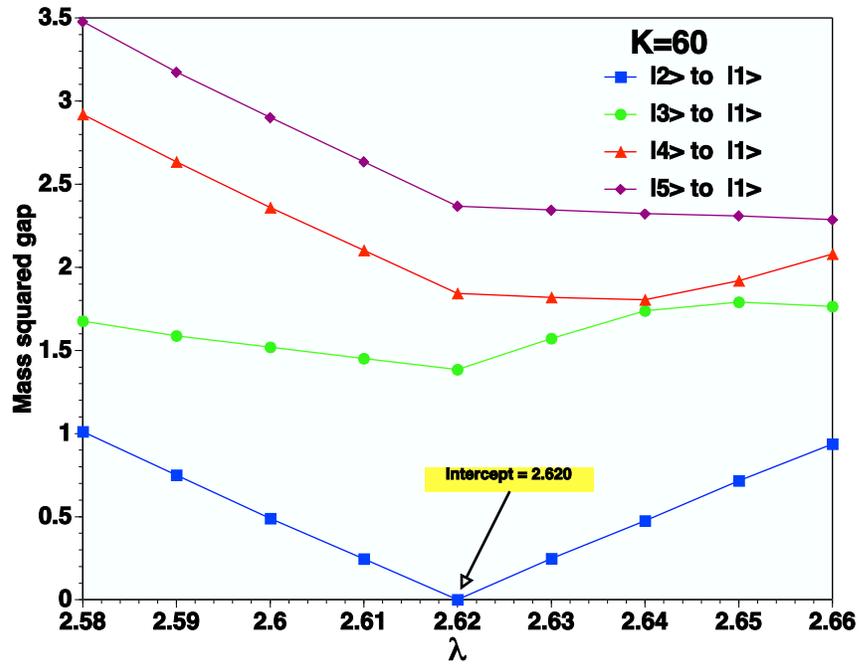}
    \caption{(Color online) Same as in Fig.
\ref{mass2gapfig1} but for $K=60$. The upper gaps are connected by
straight line segments while the straight lines of the lowest gaps are
obtained as in Fig. \ref{mass2gapfig1}(b). }
\label{mass2gapfig2}
\end{figure}

In Figs. \ref{mass2gapfig1}(b), and \ref{mass2gapfig2} we exhibit the finer 
details of level crossing for $K=55$ and $60$. Here the linearity of the
disappearing mass-squared gap is clearly observed. At low $K$ values,
we see evidence for the mixing of lowest levels (not shown in the figures)
but that mixing disappears with increasing $K$. The vanishing
of the mass-squared gap is exhibited more  clearly in Fig.
\ref{mass2gapfig2}. In this case we have observed vanishing mass-squared gap
to a very high precision (6 significant figures) which is of the same order
as the machine accuracy in single precision. i.e., there is no mixing
or the level of mixing present is at or below the level of computational
noise.

\begin{figure}[hbtp]
\centering
\includegraphics[width=5in,clip]{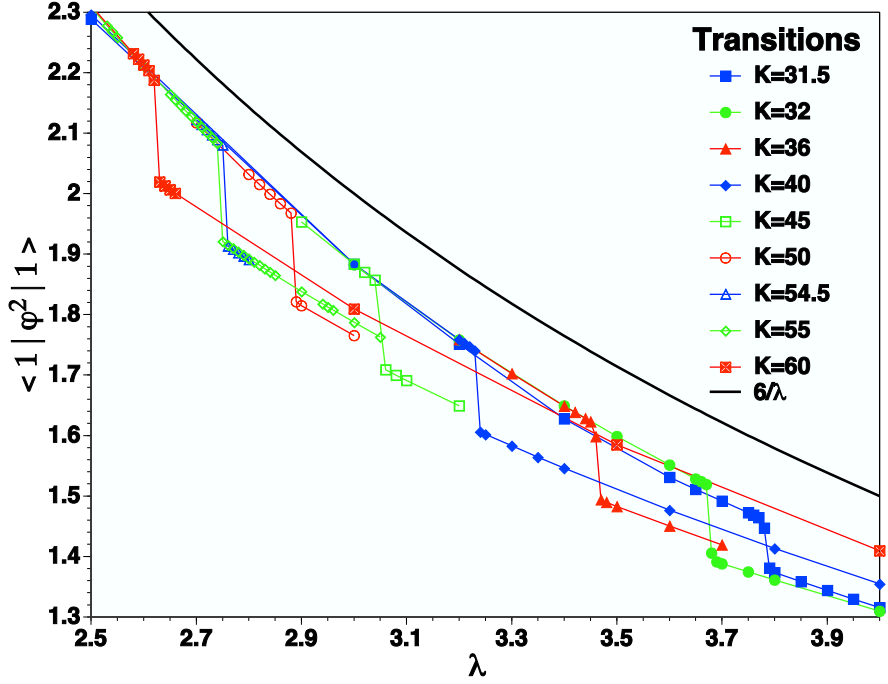}
    \caption{(Color online) $ \langle 1 \mid \phi^2 \mid 1 \rangle$ (short hand
notation for the expectation value of the integral of the normal ordered 
$ \phi^2$ operator)
as a function of $\lambda$ and selected $K$ values. For comparison we have
also shown $ \phi_{classical}^2 = 6 \frac{\mu^2}{\lambda}$ with $\mu^2=1$.  }
\label{phijump1}
\end{figure}

Another manifestation of the level crossing phenomena is found
in the expectation
value of the integral of the normal ordered $\phi^2$ operator.  
In Fig. \ref{phijump1} we show
the behavior of this observable for the lowest excitation as a function of 
 $ \lambda$ for different
values of $K$. The sudden drops match with level crossing points
of the lowest state in the 
mass-squared gap spectra. For comparison we have also shown the behavior
of $ \phi_{classical}^2 = 6 \frac{\mu^2}{\lambda}$ as a function
of $ \lambda$ in the same figure. Note that the value of $ \lambda$ at 
which the drop
occurs, $ \lambda_c$,  decreases with increasing $K$. Furthermore, we 
observe that as $K$ 
increases the difference in the $\lambda_c$ extracted from the even and the 
odd sector transitions, systematically tends towards zero. The drop in the
$\phi^2$ observable sharpens as a function of $\lambda$ with increasing
$K$. Again, this agrees well with a sharpening of level crossing
with increasing $K$.   
As discussed in next section, 
the reason for this drop is the transition of the lowest states of the
system from a kink type to a dominant kink-antikink-kink type. We will show that 
in the number density, this transition manifests as  
a  shift of 
maximal occupation from
the lowest momentum mode to the next available mode.

\begin{figure}[hbtp]
\centering
\includegraphics[width=4.5in,clip]{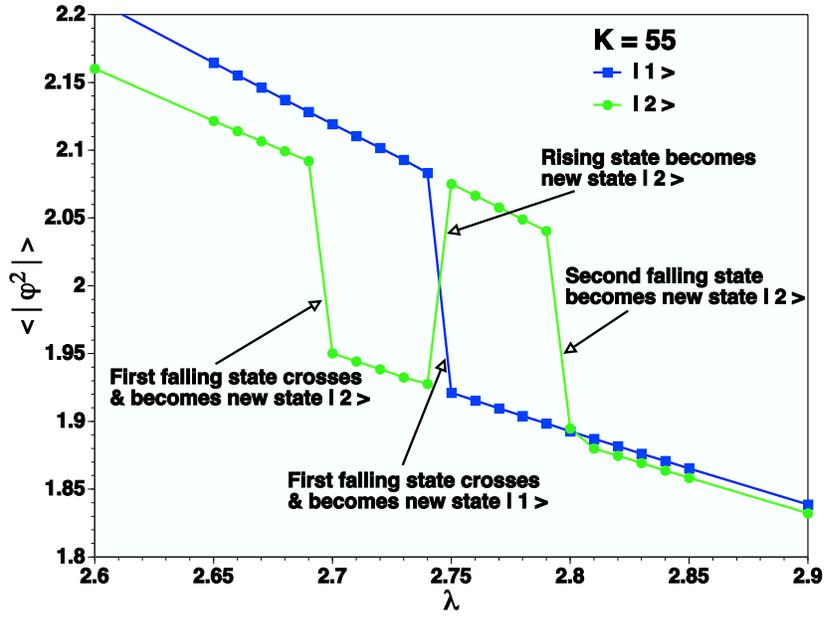}      
    \caption{(Color online) $ \langle \mid \phi^2 \mid \rangle$ (see caption to 
Fig. \ref{phijump1} for the notation)   
as a function of $\lambda$ for  $K$=55 for the lowest two excitations. }
\label{phijump12}
\end{figure}

In Fig. \ref{phijump12} we compare the
behavior of the same observable for the lowest two states for the 
even ($K=55$) particle 
sector. As one approaches the critical coupling, a set of states which 
are highly
excited states at weak coupling drop rapidly into the low-energy region. 
Between the values of the coupling, 2.65 and 2.7 the
first falling state crosses and becomes the new second state. In the
vicinity of the coupling 2.75, the first falling state crosses and 
becomes the
new lowest state and the rising state becomes the newest second state.
Close the coupling 2.8, the
second falling state becomes the new second state.     

\begin{figure}[hbtp]
\centering
\includegraphics[width=4.5in,clip]{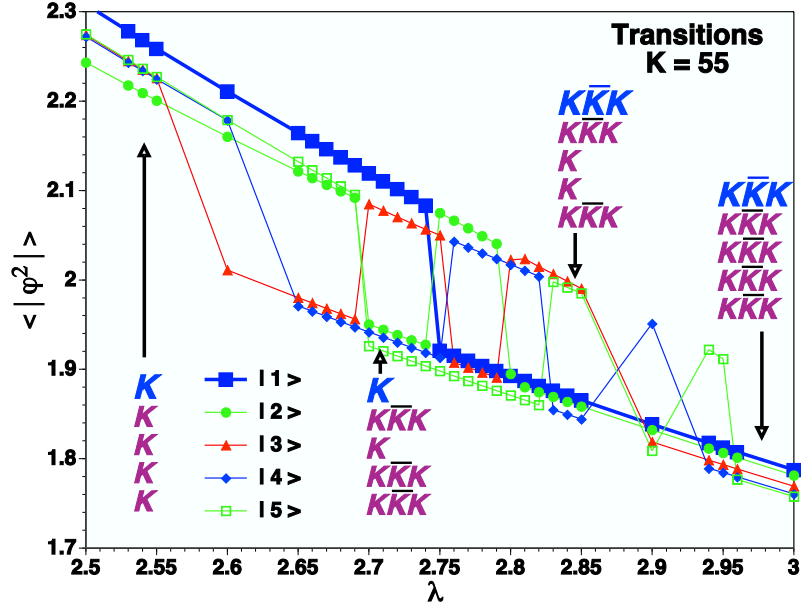}
    \caption{(Color online) $ \langle \mid \phi^2 \mid \rangle$
(see caption to
Fig. \ref{phijump1} for the notation)
as a function of $\lambda$ for  $K$=55 for the lowest five excitations. 
The pattern of transitions correspond to 5 states falling with
increasing $\lambda$ and crossing the 5 lowest states, thus replacing them
and becoming the new 5 lowest states. At selected values of $\lambda$, the
character of the  lowest states is indicated on the figure
with the top level of each column signifying the nature of the lowest
state. Successive excited states are signified by the labels proceeding
down the column. The letter $``$K" repesents $``$kink" while $``K{\bar K}K$"
represents $``$kink-antikink-kink". 
}
\label{phijump1t5}
\end{figure}

In Fig. \ref{phijump1t5} we demonstrate that
such level crossings occur for all the low-lying excitations. For
example, at the coupling of 2.5 all the lowest five excitations are
of the kink type. At the coupling of 3, all of them have
become dominantly kink-antikink-kink type. It is remarkable that in 
the entire region
of the coupling, the $\phi^2$
observable for all the low lying states cluster around that of
either a kink or a dominantly kink-antikink-kink type. i.e., we observe no
case caught in the midst of a transition. We conclude there is 
essentially no mixing
of the lowest five states of one type with any states of the other type
within numerical precision.  

All these features in our results suggest we are observing the
characteristics of a phase transition that can manifest in a system with a 
finite degrees of freedom.
\section{Calculation of other observables} 
\begin{figure}[hbtp]
\subfigure[]{
\centering
\includegraphics[width=4.5in,clip]{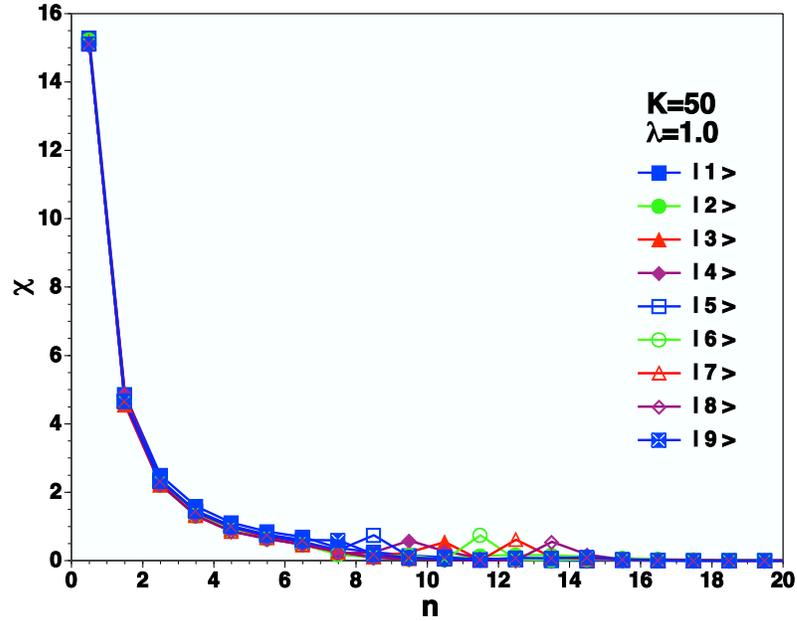}}
\subfigure[]{
\centering
\includegraphics[width=4.5in,clip]{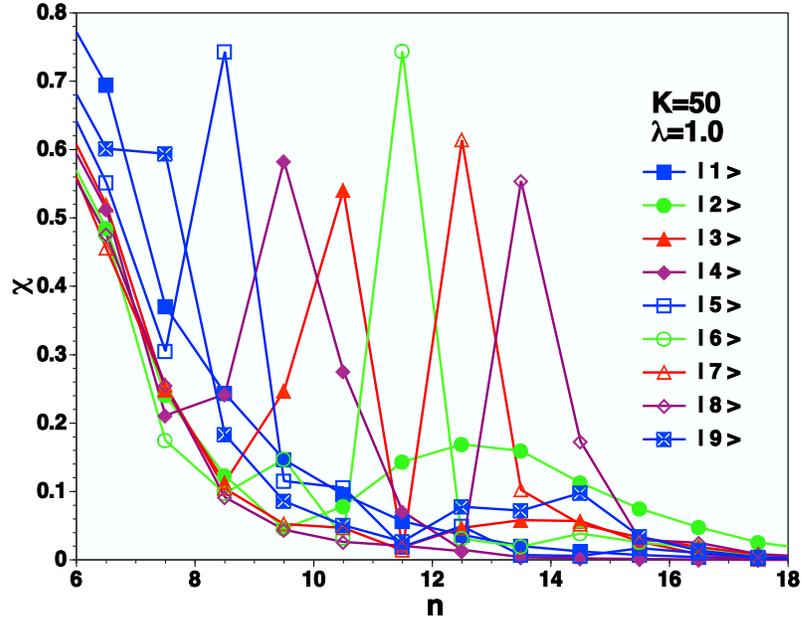}}
\caption{(Color online) (a) $\chi$ versus $n$, the half-odd integer
representing light front momentum with APBC,  for the lowest nine excitations
for $ K=50,~  \lambda=1$. (b) Same as in (a) but showing the region from
$n=6 $ to $18$ in detail.}
\label{chinsl1}
\end{figure}

To gain further understanding of the nature of the levels that cross, 
we next examine the behavior of the parton density $\chi(n)$ for the lowest
excitation for
different values of coupling. In Figs. \ref{chinsl1}(a) and \ref{chinsl1}(b)
 we present the parton
distribution function $\chi(n)$ for the lowest nine excitations at $
\lambda=1$ and $K=50$.  At small coupling, the lowest excitation is a
kink state which yields a characteristic parton distribution which 
peaks at the lowest momentum mode available. In the semi-classical
picture, the next excitation is an excited kink for which also we
expect and find a smooth distribution function. 
The excited kink, the second excitation in the spectrum, features a 
broad and smooth peak spanning $n=10.5$ to $n=17.5$ as is seen in 
Fig. \ref{chinsl1}(b).  The integrated momentum fraction under this peak accounts 
for 22\% of the momentum sum rule.

The third and higher excitations are expected to be kink plus boson 
states. The characteristic peaks in Fig. \ref{chinsl1}(b) for states 3-8 are 
observed at $n=(10.5, 9.5, 8.5, 11.5, 12.5$ and $13.5)$ respectively. 
Sharp peaks suggest a boson in a pure momentum state coupling weakly 
to the kink.  This picture is supported by the momentum fraction of 
approximately 20\% of the sum rule carried in each of these peaks, 
which corresponds well to a boson positioned around n=10 in a state 
carrying a total light-front momentum of $K=50$.

The ninth state in the spectrum does not exhibit the simple isolated 
boson plus kink features and this state may indicate the onset of 
more complicated multi-boson excitations.  It would be necessary to 
perform calculations at higher $K$ and for more states to further 
explore the higher lying features of the spectrum at this coupling.

At strong coupling the situation changes drastically. As seen from Fig.
\ref{chins} (a), at $\lambda=5 (K=55)$, the lowest 
momentum mode ($n=0.5$) for all the lowest four excitations has almost zero occupation and the 
next mode ($n=1.5$) has
maximum occupation for all four lowest excitations. 
Thus, between $\lambda=1$ and $\lambda=5$, a change in
the structure of the lowest excitations (kink) takes place for
this $K$ value.  
Can one
gain a better understanding of the parton structure of these states, in
particular the striking feature of the {\em absence} of the lowest momentum
mode? 
In the next subsection we carry out a coherent state variational 
calculation for a kink-antikink-kink state and from the calculation of
the number density confirm that the lowest excitation after the first
level crossing indeed corresponds dominantly to that of the kink-antikink-kink state.

\subsection{Coherent state variational calculation of kink-antikink-kink
state}

We have seen that a simple way to realize the parton picture of the 
classical kink solution in DLCQ is the unconstrained coherent state 
variational calculation\cite{RT}. Can one understand the 
parton structure of other topological excitations 
in the same method? In particular, we are interested in 
the kink-antikink-kink state. (For a detailed investigation of 
kink-antikink-kink dynamics in classical two dimensional $\phi^4$ theory see Ref.
\cite{mm}.)

Choose as a trial state, the coherent state
\be
\mid \alpha \rangle = {\cal N} e^{\sum_{n} \alpha_n a_n^\dagger} \mid 0
\rangle
\ee
where ${\cal N}$ is an arbitrary normalization factor.

With APBC we have
\be
\frac{\langle \alpha \mid \phi (x^-) \mid \alpha \rangle}{\langle \alpha
\mid \alpha \rangle}  =  \frac{1}{\sqrt{4 \pi}} ~ f(x^-)
\ee
with
\be
f(x^-) = \sum_{n}
\frac{1}{\sqrt{n}} \Big [ \alpha_{n}
e^{-i \frac{\pi}{L}n x^-}+ \alpha_{n}^{*}
e^{i \frac{\pi}{L}n x^-}
\Big ]~  \label{uvcf}
\ee
with $ n= \frac{1}{2}, \frac{3}{2}, ...$ .
Minimizing the expectation value of the Hamiltonian, we obtain
\be
f_{min} = \pm \sqrt{\frac{24 \pi \mu^2}{\lambda}} = \pm \sqrt{\frac{3}{g}}.
\ee

\begin{figure}[hbtp]
\subfigure[]{
\centering
\includegraphics[width=4.5in,clip]{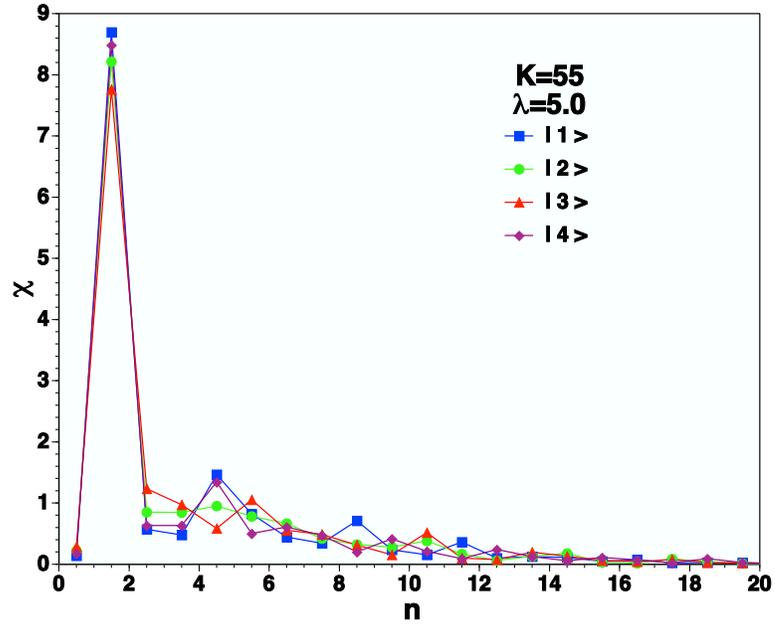}}
\subfigure[]{
\centering
\includegraphics[width=4.5in,clip]{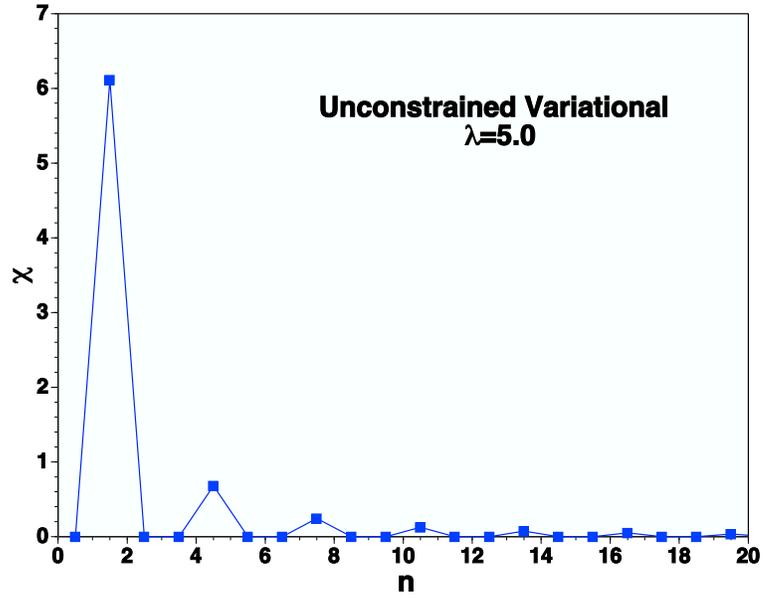}}
    \caption{(Color online) (a) $\chi$ versus $n$, the half-odd integer
representing light front momentum with APBC,  for the lowest four excitations 
for $ K=55,  ~\lambda=5$. (b) Kink-antikink-kink parton
density in unconstrained variational calculation for $\lambda$=5.}
\label{chins}
\end{figure}

Our starting point is the expression for the function $f(x^-)$ given
in Eq. (\ref{uvcf}). With $ f_{min} = \pm \sqrt{\frac{3}{g}}$, we set
\be
f(x^-) & = &+ \sqrt{3/g}, ~ ~ - L < x^ - < - 2L/3, \nonumber \\
       & = & - \sqrt{3/g}, ~ ~  -2L/3 < x^- < 0, \nonumber \\
       & = & + \sqrt{3/g}, ~ ~  0 < x^- < 2L/3, \nonumber \\
       & = & - \sqrt{3/g}, ~ ~  2L/3 < x^- < L.
\ee
This yields,
\be
\alpha_{n} =  \sqrt{3/g}~\frac{i}{\pi}~ 
\frac{1}{\sqrt{n}} ~
\Big [ 1 - 2 ~\cos (2n \pi/3) \Big ].
\ee
Thus 
\be
f(x^-) =  \frac{2}{\pi} \sqrt{3/g} \sum_{n} \frac{1}{n} \Big [ 1 -
\cos(2n\pi/3) \Big ] \sin(2n \pi x^-/L)~.
\ee
The number density of constituents, i.e., parton density
in light front theory is given by
\be
\chi(n) = \frac{\langle \alpha \mid a_n^\dagger  a_n \mid \alpha \rangle} 
{\langle \alpha \mid \alpha \rangle} = \mid \alpha_{n} \mid^2 ~.
\ee  
For $ \lambda=5$,  the 
parton density of the kink-antikink-kink state is presented in
 Fig. \ref{chins} (b). 

\begin{figure}[hbtp]
\centering
\includegraphics[width=5in,clip]{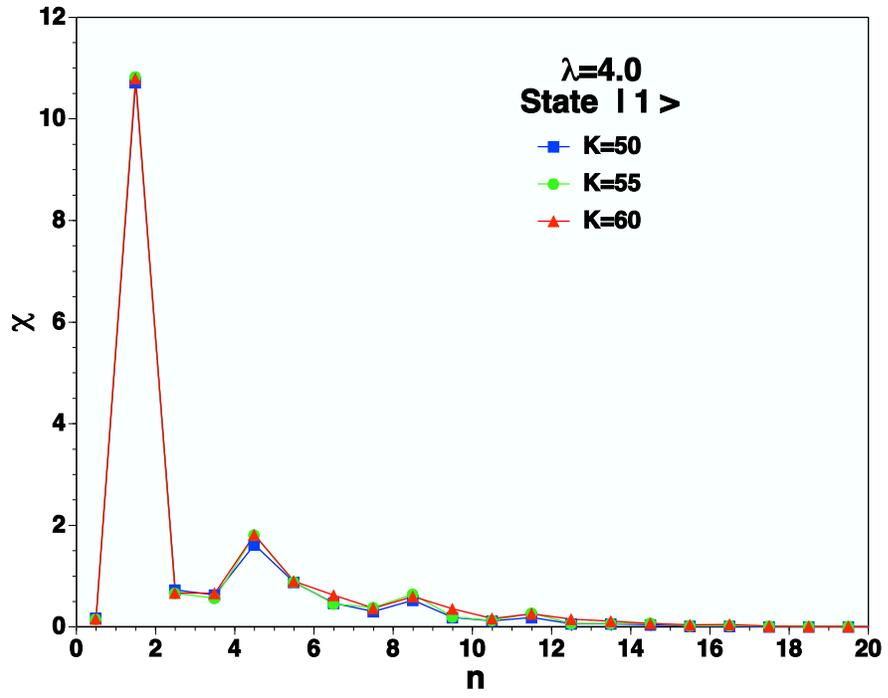}
    \caption{(Color online) $\chi$ versus $n$, the half-odd integer
representing light front momentum with APBC,  for the lowest excitation
for $ K=50, 55,$ and $60$,  $ \lambda=4$. }
\label{chinc}
\end{figure}

In Fig. \ref{chinc} we compare the number density of the lowest excitation
at $ \lambda=4$ for $K=50, 55, 60$ and show remarkable stability of results
with respect to variations in $K$. 
Further, by comparison with the lowest state in Fig. \ref{chins}(a),
we see that the shape is independent of coupling and the magnitude follows
the trend of the coherent state analysis. 


Evaluation of the number density
in the unconstrained variational calculation clearly shows that the 
lowest excitation 
after the first level crossing is definitely {\em not} a
kink but a dominantly kink-antikink-kink state. It has small admixtures
of other topological structures as is evident from the following. For a
pure kink-antikink-kink state, $\chi(n)$ vanishes for $n=\frac{5}{2}, 
\frac{7}{2}, \frac{11}{2}$ etc as
seen from Fig. \ref{chins}(b) but it doesn't
vanish for these modes for the state observed after the transition as seen
in Fig. \ref{chins}(a).  
An interesting issue is whether
we can identify the nature of the excitation
within DLCQ, i.e., without the help from unconstrained variational
calculation. This is possible since the
Hamiltonian diagonalization provides us with various eigenfuctions of 
 the lowest few excitations and
we may evaluate other observables that yield more information about the
structure of these states.  

\subsection{Fourier transform of the form factor}

\begin{figure}[hbtp]
\centering
\subfigure[]{
\includegraphics[width=4.5in,]{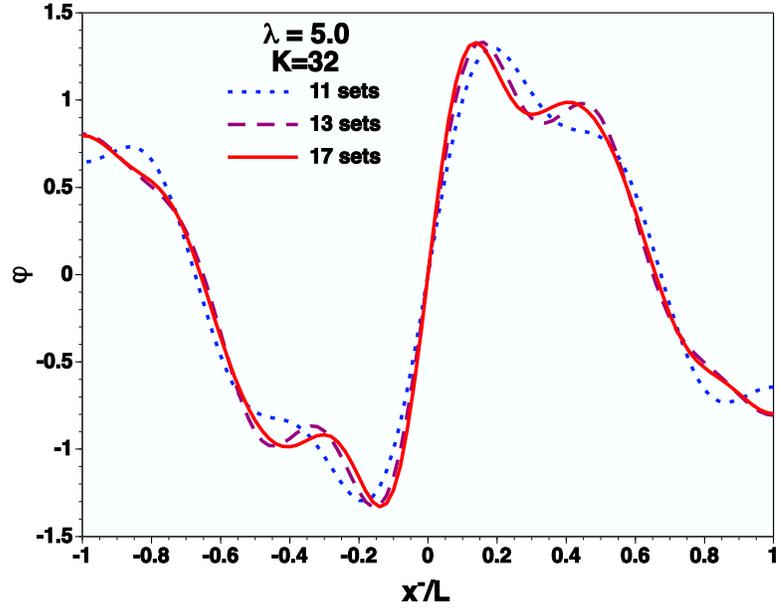}}

\subfigure[]{
\includegraphics[width=4.5in,]{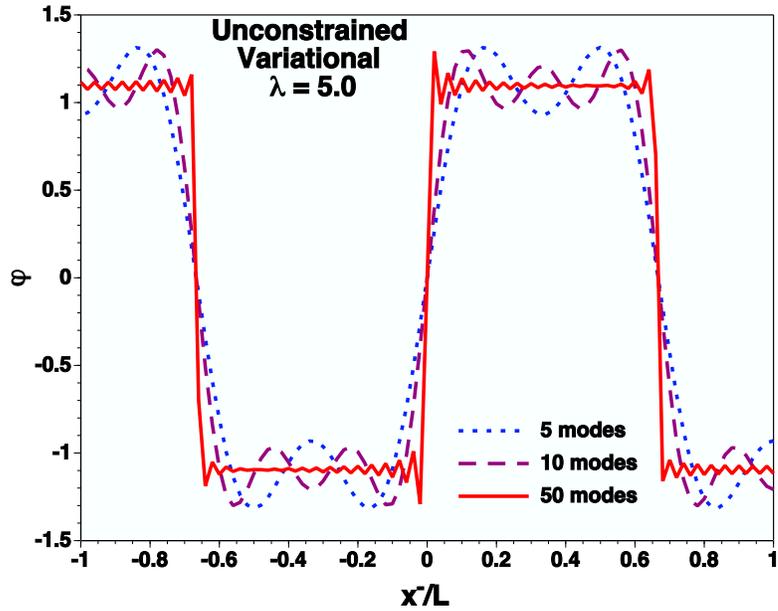}}
\caption{(Color online) (a) Fourier Transform of the kink form factor at
$\lambda$=5, $K=32$. The figure  legend indicates the number of adjoining 
momentum
transfer terms (sets) included in the summation. (b) 
Expectation value of the field operator in the kink-antikink-kink 
state  in the unconstrained variational calculation for $\lambda=5.0$.  }
\label{profile}
\end{figure}

An useful  observable that gives direct information on the nature of 
the physical state is the Fourier transform of the form 
factor of the lowest
state, which, according to Goldstone and Jackiw \cite{gj}, gives the
coordinate space profile of the topological excitation. 
The analysis of Ref. \cite{gj} was restricted to weak coupling 
theory, whereas, we are able
to perform reliable nonperturbative calculations at strong coupling.  
By explicitly calculating the form factor for different momentum \
transfers,
we perform the Fourier transform numerically. For weak coupling, the 
result is
presented in Ref. \cite{kinkplb} which reveals a kink profile.
In Fig.
\ref{profile}(a) we present the profile of the lowest excitation in DLCQ
for $ \lambda=5$ and $K=32$. Results are presented for three cases where
the magnitude of the maximum momentum transfer included in the sum for the 
Fourier transform ranges from 5 to 8. The profile is that of a
dominant kink-antikink-kink state. But as we have learned from the
evaluation of the number density, the state after the transition is not 
a pure
kink-antikink-kink configuration.  The appearance of more  structures
near the origin in Fig. \ref{profile}(a) may be a consequence of this fact.    
In Fig. \ref{profile} (b) we present the expectation value
of the field operator in the kink-antikink-kink state in the unconstrained
coherent state variational calculation. It is worthwhile to note
that the sharp features present in the profile
in the unconstrained variational calculation as the number
of modes included increases as shown in Fig. \ref{profile}(b) is an
artefact of the unconsrtained variational calculation which yields
an infinite value for the expectation value of $K$. We expect much smoother
behaviour for the constrained variational calculation which yields
finite expectation value of $K$ \cite{kinkplb}.  
Noting the close similarities of the observables in Fig. \ref{chins} (a) 
and (b) and
Fig. \ref{profile} (a) and (b) we conclude that at a critical coupling, the
lowest excitation in the topological sector with charge $\pm 1$ of 
broken phase of $\phi_2^4$ theory changes its
character from a kink state to a state dominated by kink-antikink-kink
structure.

\begin{figure}[hbtp]
\includegraphics[width=5in,clip]{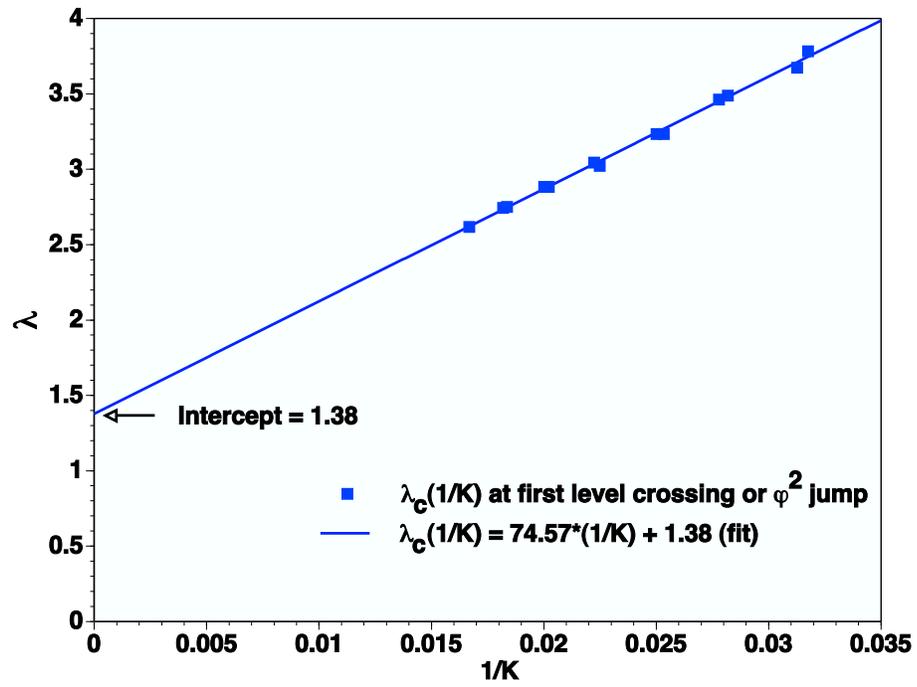}
    \caption{Critical coupling for level crossing
as a function of $\frac{1}{K}$, and an indication of the critical coupling
in the continuum limit.}
\label{1lambdac}
\end{figure}

In Fig. \ref{1lambdac} we
present the behavior of this critical coupling with $\frac{1}{K}$ 
from the analysis of vanishing mass-squared gaps and $\phi^2$ jumps  as shown in
Figs. \ref{mass2gapfig1} and \ref{phijump1} and
extract its value in the continuum limit as $\lambda_c=1.38$.   

We have already observed that the mass-squared gap vanishes linearly
at the critical coupling for level crossing. This implies that, 
parameterizing the behavior
of the mass-squared gap near the critical coupling as 
$ \delta M^2 \approx
(\lambda_c  - \lambda)^{\nu}$, we have obtained the exponent $ \nu=1.0$.

\section{Discussion, Summary and Conclusions}
To investigate the strong coupling region of the topological sector of 
two dimensional $\phi^4$ theory we have utilized DLCQ which provides us
with all the advantages of the Hamiltonian approach with additional features
of light front quantization.. 
A major bonus of using DLCQ is the detailed information we gain about 
the parton structure of the states. We have shown that between
$\lambda=1$ and $\lambda=2$, level crossings 
occur in the continuum limit. 
The important issues to resolve are the nature of this
transition and its physical implications. 
To settle 
this issue we have studied the expectation value of the integral
of the normal ordered 
$\phi^2$ operator in the lowest excitations of the system. 
We have observed a sharp  drop in this observable. 
We also observed a  corresponding change in behavior of 
the number density in the lowest excitations, namely, the shift of the
maximum occupation from the lowest momentum mode to the next higher
momentum mode. 

In the weak coupling region, one can use analytical variational
calculations with a coherent state ansatz for the lowest state
to gain physical insights for the DLCQ data. Following
Rozowsky and Thorn \cite{RT} we have carried out 
unconstrained variational calculations but with APBC. 
Variational calculations predict maximum occupation in 
the lowest momentum mode for the lowest eigenstate. Our numerical results
show that this simple semi-classical 
picture becomes invalid as the coupling grows greater than 1. A coherent state
variational calculation corresponding to kink-antikink-kink state predicts
maximum occupation for the next higher momentum mode and almost zero
occupation of the lowest momentum mode.
This is consistent with the observed properties of our 
states above the transition.

In Ref. \cite{kinkplb}, following Goldstone and Jackiw, we have 
calculated the Fourier transform of
the form factor of the kink state in DLCQ at weak coupling and demonstrated
consistency with the classical solution. At strong coupling after the drop
in the $\phi^2$ observable, we have shown here that the profile 
calculated in DLCQ is
approximately that of a kink-antikink-kink state.
From the analysis of vanishing mass-squared gap and the drop in the
$\phi^2$ observable we have extracted the critical
coupling for the first transition in the
infinite volume limit as  $\lambda_c = 1.38$. 

The transition that we have observed is very sensitive to the boundary
conditions. We have observed that in similar calculations performed with 
PBC such a transition
is absent at the same value of $\lambda$.  With PBC, we do observe level
crossings at much higher coupling which appears to correspond to the 
transition from a kink-antikink state to a mixture of states with a dominant 
kink-antikink-kink-antikink component. 

In the two-dimensional Ising model it is well-known that the physical 
mechanism
for the symmetry restoring phase transition is the phenomena of kink
condensation \cite{sf,kogut}.  
It is known that at strong coupling, the $\phi_2^4$ theory undergoes a 
symmetry
restoring phase transition. 
As far as we know, the physical mechanism behind the phase transition 
has not been investigated before in the two dimensional $\phi_2^4 $ 
quantum field theory. 
We have demonstrated that in this
theory, at strong coupling, it is energetically favourable for a
dominantly kink-antikink-kink configuration to be the lowest excitation rather 
than a kink configuration. At still higher coupling we have
observed additional level crossings for the lowest state for both PBC and APBC. Further
investigations are necessary to clarify the nature of the lowest
excitation after these transitions and to quantify the critical point
of the transition. In the light of all our observations,  
we interpret the observed level crossing presented here as the onset of
kink condensation which leads to the restoration of symmetry. 
 
\acknowledgments
This work is supported in part by the Indo-US
Collaboration project jointly funded by the U.S. National Science
Foundation (INT0137066) and the Department of Science and Technology, India
(DST/INT/US (NSF-RP075)/2001). This work is also supported in part by the US
Department of Energy, Grant No. DE-FG02-87ER40371, Division of
High Energy and Nuclear Physics.  


\eject



\eject    

\eject

\eject

\eject

\end{document}